\documentclass[sigconf,9pt]{acmart}
\AtBeginDocument{%
  }

\setcopyright{acmlicensed}
\copyrightyear{2026}
\acmYear{2026}
\setcopyright{cc}
\setcctype{by-nc-nd}
\acmConference[SIGMOD Companion '26]{Companion of the International Conference on Management of Data}{May 31-June 05, 2026}{Bengaluru, India}
\acmBooktitle{Companion of the International Conference on Management of Data (SIGMOD Companion '26), May 31-June 05, 2026, Bengaluru, India}
\acmDOI{10.1145/3788853.3801609}
\acmISBN{979-8-4007-2450-3/2026/05}

\usepackage{enumitem}
\usepackage{multirow}
\usepackage{makecell}
\usepackage{mathtools}
\usepackage{setspace}
\usepackage[linesnumbered,ruled,noend]{algorithm2e}
\usepackage{xcolor}
\usepackage{framed}
\usepackage{multicol}
\usepackage{balance}
\usepackage[utf8]{inputenc}
\usepackage{amsthm}
\usepackage{amsmath}

\begin{document}

\title{AkasicDB: Demonstrating Omni RAG with a Unified Vector-Graph-Relational DBMS}

\author{Geonho Lee}
\email{ghlee5084@kaist.ac.kr}
\affiliation{%
  \institution{KAIST}
  \city{Daejeon}
  \country{Republic of Korea}
}

\author{Jeongho Park}
\email{jhpark@graphai.io}
\affiliation{%
  \institution{GraphAI}
  \city{Daejeon}
  \country{Republic of Korea}
}

\author{Donghyoung Han}
\email{dhhan@graphai.io}
\affiliation{%
  \institution{GraphAI}
  \city{Daejeon}
  \country{Republic of Korea}
}

\author{Min-Soo Kim}
\authornote{Corresponding author.}
\email{minsoo.k@kaist.ac.kr}
\affiliation{%
  \institution{KAIST}
  \city{Daejeon}
  \country{Republic of Korea}
}
\renewcommand{\shortauthors}{Geonho Lee, Jeongho Park, Donghyoung Han, \& Min-Soo Kim}

\begin{abstract}
Recent Retrieval-Augmented Generation (RAG) systems increasingly combine vector retrieval with structured knowledge, such as Graph RAG and Filtered vector search.
However, existing database architectures struggle to support such complex RAG workflows efficiently, as they rely on out-of-DB pipelines or in-DB non-native integration, leading to high overhead.
This demo paper presents AkasicDB, a database system that natively supports such RAG workflows by jointly executing vector similarity search, graph traversal, and relational filtering within a single execution framework.
AkasicDB extends our prior work, Chimera, with native vector support to enable such unified execution.
Based on AkasicDB, we demonstrate the first native integration of Vector–Graph–Relational RAG, which we refer to as \emph{Omni RAG}.
Through an interactive chat-style demonstration, users execute and visualize Omni RAG queries, directly experiencing its superior retrieval and reasoning over vector-only approaches while observing the practical limitations of existing database architectures in supporting Omni RAG.
A demonstration video is available at \url{https://youtu.be/8d09_dtrEIM}
\end{abstract}





\maketitle
\section{Introduction}

Retrieval-Augmented Generation (RAG) has emerged as a key paradigm for grounding Large Language Models (LLMs) with external knowledge to improve factuality and reasoning.
In practice, RAG has predominantly been realized through vector-centric retrieval pipelines, where unstructured text is retrieved via embedding similarity and passed to a language model.
Building on this paradigm, recent studies have investigated Graph RAG~\cite{zhou2026graphrag, guo2025lightrag}, which leverages graph traversal to improve multi-hop reasoning, as well as Filtered vector search~\cite{zhang2023vbase}, where relational predicates are applied to constrain vector retrieval for improved factual grounding.
Despite these advances, Graph RAG and Filtered vector search have largely evolved in isolation, resulting in RAG pipelines that combine vector similarity search with either graph traversal or relational filtering~\cite{li2025neutronrag}, but no single workflow has demonstrated the joint integration of all three.
This limitation primarily stems from the absence of a suitable database system architecture that can \emph{natively integrate} these retrieval modalities within one execution framework.
As a result, attempts to combine them typically rely on either orchestrating multiple native systems outside the DBMS or stitching the modalities together inside the DBMS through non-native extensions, which hinders unified execution, visualization, and interaction of such RAG workflows as a cohesive whole.

To address this architectural limitation, we leverage our prior work, Chimera~\cite{lee2024chimera}, which introduced a unified query processing for graph and relational data.
Building on this foundation, we extend Chimera with native vector support and develop AkasicDB, a database system that enables vector similarity search to be performed within the same execution framework together with graph traversal and relational filtering for RAG workflows.

Based on AkasicDB, we demonstrate the first native integration of Vector–Graph–Relational RAG, which we refer to as Omni RAG. 
The notion of native integration is explained in detail in Section~\ref{sec:system}.
In the demonstration, users interact with a web-based interface that visualizes retrieved vector, graph, and relational data, along with a chat-style interface that runs the RAG pipeline.
Through this demonstration, users can directly experience that Omni RAG provides more effective retrieval and reasoning than vector-only approaches.
We further contrast this experience with representative \emph{out-of-DB integration of native systems} (e.g., Neo4j, Milvus) as well as \emph{in-DB non-native integration} based on PostgreSQL with graph and vector extensions (e.g., Apache AGE, pgvector), highlighting their practical limitations in supporting Omni RAG.

\vspace{-1em}
\section{Omni RAG}
\label{sec:omnirag}

\begin{figure}[hbtp]
    \centerline{\includegraphics[width=3.4in]{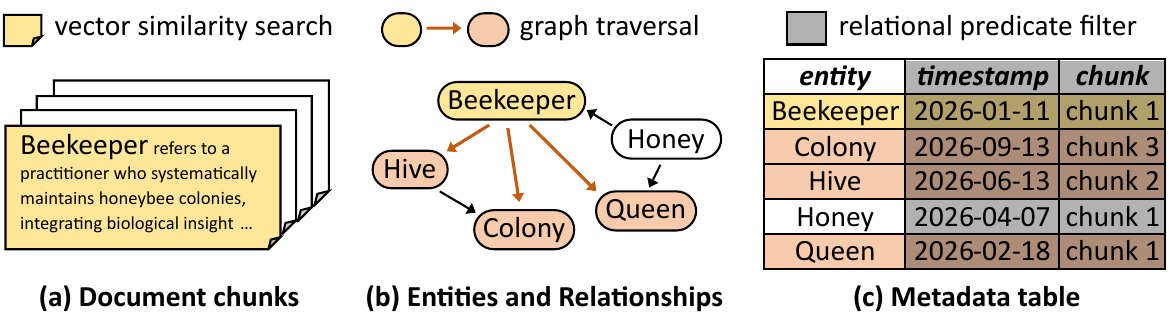}}
    \vspace{-4mm}
    \caption{Example of Vector-Graph-Relational RAG.}
    \label{fig:omnirag}
\end{figure}

Vector RAG embeds both user questions and document chunks into a shared vector space and retrieves based on embedding vector similarity. 
Figure~\ref{fig:omnirag}(a) illustrates vector RAG using the example question “What are the essential skills and knowledge needed for new beekeepers to succeed?”. 
In this setting, the question embedding is used to retrieve a set of top-$k$ document chunks that are semantically similar, which are then provided as context for generation. 
Vector RAG has proven effective for leveraging unstructured textual knowledge at scale, and thus serves as the default retrieval paradigm in many contemporary RAG systems.

Meanwhile, graph RAG augments document chunks with an explicitly constructed knowledge graph, as illustrated in Figure~\ref{fig:omnirag}(b).
Rather than retrieving isolated text chunks, Graph RAG aims to provide richer and more comprehensive context by traversing entity relationships, such as those among Beekeeper, Hive, Colony, and Queen.
Filtered vector search, illustrated in Figure~\ref{fig:omnirag}(c), leverages structured metadata (e.g., timestamp or entity type) to constrain the search space before or during vector retrieval.
While Graph RAG focuses on expanding the scope of retrieved information through relational structure, filtered vector search emphasizes excluding irrelevant content to improve precision.

Building on these complementary retrieval paradigms, we define \emph{Omni RAG} as a RAG formulation that jointly accesses vector, graph, and relational data, and explicitly specifies what to retrieve from each modality.
In Omni RAG, retrieval combines vector similarity search over unstructured text, graph traversal over entity relationships, and relational filtering over structured attributes.

\vspace{-0.5em}
\vspace{-1.75em}
\section{AkasicDB System Overview}
\label{sec:system}

This section discusses why existing architectures make Omni RAG difficult in practice and how AkasicDB overcomes these limitations.

\subsection{System Design for Omni RAG}

\begin{figure}[hbtp]
    \centerline{\includegraphics[width=3.4in]{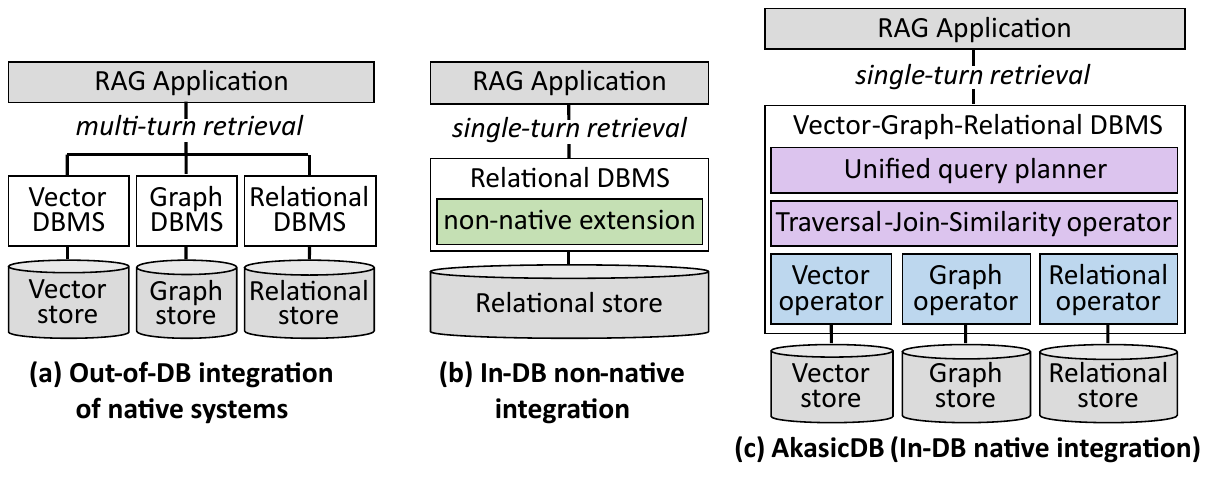}}
    \vspace{-4mm}
    \caption{Comparison of architectures.}
    \label{fig:architecture}
\end{figure}

A straightforward way to implement Omni RAG is to rely on multiple specialized systems, as illustrated in Figure~\ref{fig:architecture}(a).
In this setting, a RAG application queries native vector, graph, and relational DBMSs separately and integrates their results outside the database---\emph{out-of-DB integration of native systems}.
Although this approach is functionally feasible, it incurs high operational costs to manage multiple systems and often leads to significant inefficiencies.
Since the context length of the LLM is limited, RAG ultimately requires only the top-$k$ most relevant results.
However, when retrieval is performed across multiple systems, it is carried out in a multi-turn manner, making it difficult to determine how many results each system should produce.
For example, the selectivity of graph traversal or relational filtering is not known in advance, and thus each system must conservatively generate more intermediate results than actually needed.
These intermediate results are materialized, transferred to the application layer, and only then pruned by a final top-$k$ restriction.
As a result, this process often produces excessive intermediate data, leading to redundant data transfer, increased latency, and inefficient resource utilization.

To mitigate these issues, it is desirable to process Omni RAG through an integrated system.
In such a setting, retrieval across graph, relational, and vector data can be expressed as a single SQL extended query.
Figure~\ref{fig:omnirag-query} shows an example query, inspired by LightRAG~\cite{guo2025lightrag} and extended with relational filtering, that jointly retrieves over the three modalities followed by a final \texttt{LIMIT} $k$.
Here, SQL/PGQ~\cite{deutsch2022graph} extends SQL for graph pattern matching, while the \texttt{<->} operator denotes vector similarity.
This formulation enables the query to be planned and executed in a single turn, allowing the top-$k$ constraint to be enforced during execution and avoiding excessive intermediate results.

\vspace{0.5em}
\setlength{\FrameSep}{5pt}
\begin{center}
\begin{minipage}{0.95\linewidth}
\begin{framed}
\noindent\texttt{\textbf{SELECT} chunk} \\[-1.5pt]
\noindent\texttt{\textbf{FROM GRAPH\_TABLE} (} \\[-1.5pt]
\noindent\texttt{\hspace*{1.5em}\textbf{MATCH} (src:entity)-[:related\_to]->(dst:entity)} \\[-1.5pt]
\noindent\texttt{\hspace*{1.5em}\textbf{COLUMNS} (} \\[-1.5pt]
\noindent\texttt{\hspace*{3em}src.embedding \textbf{AS} src\_embedding,} \\[-1.5pt]
\noindent\texttt{\hspace*{3em}dst.chunk \textbf{AS} chunk,} \\[-1.5pt]
\noindent\texttt{\hspace*{3em}dst.timestamp \textbf{AS} timestamp} \\[-1.5pt]
\noindent\texttt{\hspace*{1.5em})} \\[-1.5pt]
\noindent\texttt{) \textbf{WHERE} timestamp \textbf{IN} :selected\_time\_range} \\[-1.5pt]
\noindent\texttt{\textbf{ORDER BY} src\_embedding <-> :question\_embedding} \\[-1.5pt]
\noindent\texttt{\textbf{LIMIT} 5;} 
\end{framed}
\vspace{-1.2em}
\captionof{figure}{An SQL extended query example for Omni RAG.}
\label{fig:omnirag-query}
\end{minipage}
\end{center}
\vspace{0.5em}

Figure~\ref{fig:architecture}(b) illustrates an extension-based approach that enables a single-query formulation of Omni RAG---\emph{in-DB non-native integration}.
Highly extensible DBMSs support this approach by allowing graph querying and vector similarity search to be added through extensions.
However, these capabilities remain non-native to the core engine.
In practice, graph traversal is translated into sequences of relational joins, and vector data are typically stored and processed as generic array types.
As a result, both graph traversal and vector search incur substantial overhead, and query optimization remains dominated by a relational planner that lacks awareness of graph- and vector-specific execution characteristics.
Consequently, although Figure~\ref{fig:architecture}(b) supports the expression of Omni RAG queries, it still suffers from inefficient execution.

To address these limitations, we adopt the architecture shown in Figure~\ref{fig:architecture}(c) and develop AkasicDB, which realizes \emph{in-DB native integration} for Omni RAG.
AkasicDB consists of (i) dedicated storage components for vector, graph, and relational data, and (ii) a unified query processing layer that coordinates traversal, join, and similarity search within the same execution framework.
This architecture provides the foundation for efficient Omni RAG processing, which we detail in the following subsections.

\vspace{-1em}
\vspace{-1.5em}

\begin{figure*}[hbtp]
    \centerline{\includegraphics[width=6.4in]{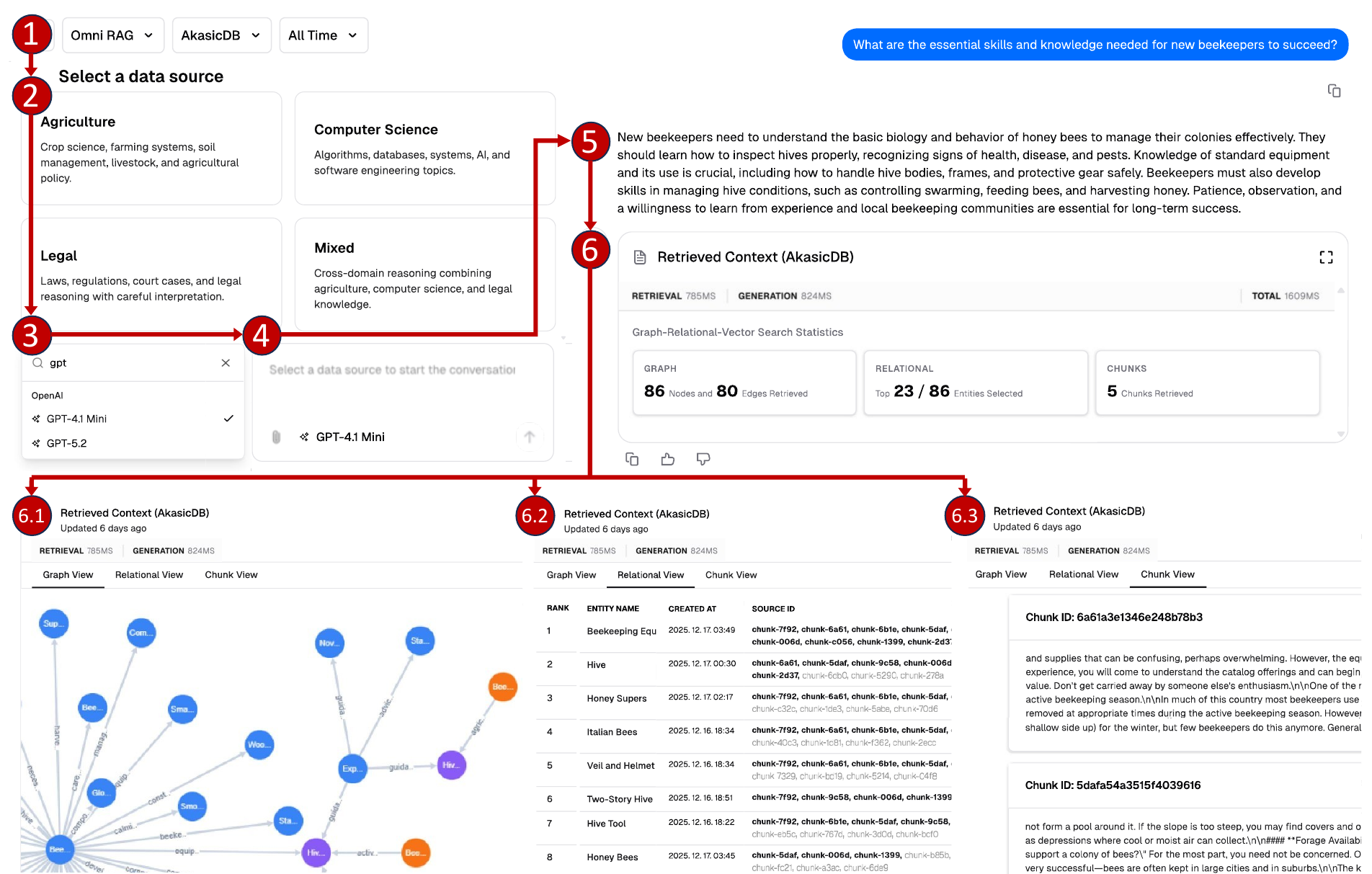}}
    \vspace{-1.0mm}
    \caption{End-to-end demonstration workflow of Omni RAG.}
    \label{fig:demonstration}
\vspace{-1.2em}
\end{figure*}

\subsection{Vector-Graph-Relational Storage}

AkasicDB extends the dual-store architecture of Chimera into a triple-store design by incorporating a native vector store alongside the graph and relational stores. 
The graph store maintains graph topology using adjacency-list, while the relational store is built on PostgreSQL tables. 
Vectors are stored separately in a dedicated disk-based storage called segments to support efficient similarity search.
These three stores are co-located within a single system and share a common transaction manager, ensuring consistent and up-to-date access across graph, relational, and vector data.

AkasicDB provides indexing mechanisms for each store according to its data representation. 
Both the graph store and the relational store support B-tree indexes to accelerate attribute-based filtering and access to graph elements. 
For vector data, AkasicDB supports multiple indexes, including HNSW, IVF, and Vamana, which perform approximate nearest neighbor search to efficiently retrieve similar vectors at scale.

\subsection{Unified Query Processing}

AkasicDB builds on Chimera’s Traversal-Join (TJ) operator, which enables graph traversal and relational joins to be executed within a single query plan by taking both topology and property operands as input. 
To support Omni RAG workloads, AkasicDB extends this operator by integrating vector similarity search following the VBASE~\cite{zhang2023vbase} design. 
Specifically, AkasicDB exposes ANN index traversal as an iterator-based operator with an \texttt{Open/Next/Close} interface, allowing similarity search to be embedded into a Volcano-style execution framework. 
Under this design, all vector, graph, and relational operations incrementally produces records via \texttt{Next}, while global top-$k$ semantics are enforced by upstream \texttt{ORDER BY} and \texttt{LIMIT} operators.
As a result, operations across the three modalities are seamlessly composed within a single execution plan, enabling unified and efficient query processing for Omni RAG.

\section{Demonstration Scenarios}
\label{sec:demonstration}

\subsection{End-to-End Experience}

We demonstrate Omni RAG in six steps, as illustrated in Figure~\ref{fig:demonstration}.

\noindent\textbf{Step 1. Configuration.}
The demonstration begins by selecting the RAG mode and target system. 
Depending on this selection, users are guided into different demonstration scenarios: one focusing on comparing Vector RAG and Omni RAG, and another focusing on experiencing performance differences across systems.

\noindent\textbf{Step 2. Data Source Selection.}
Users select a knowledge domain (e.g., Agriculture, Computer Science, Legal, or Mixed), enabling cross-domain Omni RAG over vector-graph-relational data.

\noindent\textbf{Step 3. Model Selection.}
Users then select an LLM for answer generation from a list of supported models, allowing seamless switching between models within the same retrieval pipeline.

\noindent\textbf{Step 4. Input Question.}
Users either select a suggested question or freely input their own natural language question. 
In the running example, the user asks: “What are the essential skills and knowledge needed for new beekeepers to succeed?”.

\noindent\textbf{Step 5. LLM Response.}
Based on the selected configuration, the system retrieves relevant context and presents the generated answer in a conversational chat interface using the chosen LLM.

\noindent\textbf{Step 6. Retrieval Statistics and Inspection.}
Users inspect detailed retrieval statistics, including retrieval counts across modalities such as the number of vertices and edges traversed, relational entities selected, and text chunks retrieved.
The interface reports retrieval time and generation time separately.
Users can further expand this view to examine modality-specific results in detail:

\begin{description}[labelindent=1em,labelsep=0.4em,leftmargin=*,itemsep=0.2em]

\item[Step 6.1. Graph view.] visualizes the retrieved subgraph as an interactive Cytoscape layout, where vertex colors distinguish different entity roles and edges represent retrieved relationships.

\item[Step 6.2. Relational view.] shows retrieved entities in a tabular layout, with relevant records highlighted to indicate entities that contributed to the retrieved text chunks.

\item[Step 6.3. Chunk view.] displays the retrieved text chunks in a document-style layout, allowing users to directly inspect the textual evidence used for generation.

\end{description}
\vspace{-0.5em}

\subsection{Scenario 1 -- Answer Quality Comparison}

In this scenario, users configure the system to generate answers using both Vector RAG and Omni RAG for the same question and data source.
The interface presents the resulting answers side by side, allowing users to directly compare their quality and vote which response better.
Alongside human voting, we also assess the same answers using an LLM-as-a-judge, which has been shown to correlate well with human judgments in natural language generation evaluation~\cite{wang2023chatgpt, zheng2023judging}.
This enables users to observe when automated evaluation agrees with or diverges from human preferences.
Through this interaction, the scenario highlights how Omni RAG produces more informative and useful responses and engages users in understanding both answer quality and evaluation practices.

\vspace{-0.5em}
\subsection{Scenario 2 -- System Efficiency Comparison}

In this scenario, users vary the backend system and adjust the date range to control retrieval selectivity.
The interface reports retrieval time and generation time separately, allowing direct observation of how different system architectures respond.
By issuing the same question under different configurations, users can experience how retrieval efficiency improves chat responsiveness, highlighting the impact of database system architecture on RAG performance.
\section{Experiments}

\noindent\textbf{Settings.}
We evaluate Omni RAG queries on four datasets derived from the UltraDomain~\cite{guo2025lightrag} benchmark: Agriculture (2.0M tokens), CS (2.3M tokens), Legal (5.1M tokens), and Mix (0.6M tokens). 
We compare AkasicDB against three alternative system configurations.
PGVector+AGE denotes an in-DB non-native integration based on PostgreSQL, where vector and graph functionalities are provided via extensions (pgvector and Apache AGE).
Neo4j+Milvus represents an out-of-DB integration of native systems, where graph data are managed by Neo4j, relational data are stored as key--value properties in Neo4j, and vector data are handled by Milvus.
In this configuration, we use a fixed multi-turn retrieval that performs vector search in Milvus first (to leverage its vector index), followed by graph traversal and relational filtering in Neo4j.
Neo4j Vector denotes an in-DB non-native integration that uses a fixed query plan, executing graph traversal and filtering in Neo4j first and then performing vector search via vector extension.
We measure latency using time-to-first-token (TTFT), which includes both retrieval and generation time.
All experiments were conducted on a single server with two Intel Xeon Gold 6326 CPUs and 1024\,GB RAM.

\noindent\textbf{Answer Quality Comparison.}
Table~\ref{tab:quality_latency_grouped} (upper part) reports the answer quality of Vector RAG and Omni RAG using the winning ratio evaluated by an LLM-as-a-judge following ~\cite{zhou2026graphrag}.
Across all datasets, Omni RAG consistently outperforms Vector RAG by 4–28 percentage points, with the largest gain on Mix.
This indicates that jointly leveraging graph, relational, and vector information enables more accurate answers than vector-only retrieval.

\noindent\textbf{System Efficiency Comparison.}
Table~\ref{tab:quality_latency_grouped} (lower part) presents the system efficiency of Omni RAG in terms of latency under a selectivity setting of 20\%.
AkasicDB achieves the lowest retrieval time across all datasets, reducing latency by orders of magnitude compared to PGVector+AGE and by 2–9$\times$ compared to Neo4j and Milvus configuration.
Generation time, which is dominated by LLM inference and thus incurs substantial latency, is reported for completeness; however, under complex Omni RAG retrieval settings, retrieval time can become a primary bottleneck.

\begin{table}[t]
    \centering
    \caption{Winning Ratio (\%) and Latency (msec) of Omni RAG.}
    \vspace{-1em}
    \label{tab:quality_latency_grouped}
    \renewcommand{\arraystretch}{0.70}
    \setlength{\tabcolsep}{1.2pt}
    \begin{tabular}{llcccc}
        \toprule
        \multirow{2}{*}{Metric} 
        & \multirow{2}{*}{Configuration} 
        & \multicolumn{4}{c}{Datasets} \\
        \cmidrule(lr){3-6}
        & & Agriculture & CS & Legal & Mix \\
        \midrule

        \multirow{2}{*}{Winning Ratio} 
        & Vector RAG 
        & 45\% & 48\% & 47\% & 36\% \\ \cmidrule{2-6}
        & Omni RAG 
        & \textbf{\underline{55\%}} & \textbf{\underline{52\%}} & \textbf{\underline{53\%}} & \textbf{\underline{64\%}} \\
        \midrule\midrule

        \multirow{4}{*}{\makecell[l]{Retrieval Time \\for Omni RAG}} 
        & PGVector+AGE 
        & 18,573 & 21,278 & 11,079 & 4,728 \\ \cmidrule{2-6}
        & Milvus+Neo4j 
        & 7,099 & 3,001 & 4,783 & 1,536 \\ \cmidrule{2-6}
        & Neo4j Vector 
        & 4,042 & 4,497 & 2,409 & 1,323 \\ \cmidrule{2-6}
        & AkasicDB 
        & \textbf{\underline{802}} & \textbf{\underline{942}} & \textbf{\underline{697}} & \textbf{\underline{583}} \\
        \midrule\midrule

        \makecell[l]{Generation Time \\for Omni RAG} 
        & gpt-4.1-mini
        & 883 & 876 & 882 & 881 \\
        \bottomrule
    \end{tabular}
\vspace{-1.5em}
\end{table}

\section{Conclusions}

This demonstration presented AkasicDB, a vector-graph-relational DBMS that enables efficient Omni RAG within a single system. 
We showed that existing architectures make Omni RAG difficult to realize in practice, and that AkasicDB addresses these challenges. 
Through interactive scenarios, users can compare the quality of Vector RAG and Omni RAG responses and directly experience performance differences across systems. 
Overall, this demonstration highlights the practical benefits of unified vector-graph-relational processing for emerging RAG workloads.


\bibliographystyle{ACM-Reference-Format}
\bibliography{main}


\end{document}